# Optimal Partitioned-Interval Detection Binary Quantum Receiver with Practical Devices

Ke Li, Bing Zhu
Department of Electronic Engineering and Information Science,
University of Science and Technology of China, Hefei, Anhui 230027, China
e-mail: zbing@ustc.edu.cn

**Abstract**

Partitioned-interval detection binary quantum receiver with non-ideal devices is theoretically analyzed. Using global optimized partition strategy, relatively large gain over standard quantum limit (SQL) is obtained with small partition number for certain mean photon number.

## I. INTRODUCTION

In 1970s, based on a full quantum analysis, Helstrom obtained the ultimate lower bound to the error probability of hypothesis test [1]. Since then, many efforts have been devoted to design practical receivers able to approach such a bound [2-8]. Recently, based on Dolinar receiver [2,3] and optimal displacement receiver [4,5], Vilnrotter proposed a new binary quantum receiver with partitioned-interval detection and constant-intensity local lasers [6]. Compared with Dolinar receiver, it is easier to implement and more suitable for high rate operation. At the same time, in ideal case, this partitioned receiver can obtain relatively big gain over optimal displacement receiver (ODR) and it can fill the performance gap between ODR and Helstrom bound to some extent. But there is no analysis about Vilnrotter's receiver in non-ideal conditions.

In this paper, an analytical model of the non-ideal partitioned receiver is discussed and simulated.

## II. ANALYTICAL MODEL WITH NON-IDEAL FACTORS

For BPSK modulation, the coherent state signal $|-\alpha\rangle$ and $|\alpha\rangle$ with *a priori* probability $p_0$ and $p_1$ are received, respectively corresponding to hypothesis $H_0$ and $H_1$. Fig. 1 depicts the signal model of partitioned receiver.

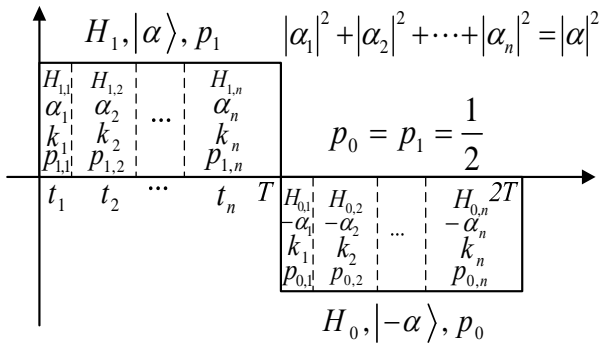

Fig. 1. The signal model of partitioned receiver.

Partitioned receiver partitions the signal interval to several segments, and performs optimal displacement measurements on each segment. In Fig. 1, the signal interval $T$ is partitioned to $N$ disjoint segments $t_1, t_2, \cdots, t_N$ ($t_1+t_2+\cdots+t_N = T$). Each segment corresponds to a coherent state signal $|\pm\alpha_i\rangle$ ($i=1,2,...,N$), which satisfy $|\alpha_1|^2 + |\alpha_2|^2 + \cdots + |\alpha_N|^2 = |\alpha|^2$. And $k_i$ ($i=1,2,...,N$) is the photon counts of each segment. Considering the "modified sequence" interpretation [6], we can obtain iterative equations (1) - (4).

$$\begin{cases} p_{1,i} = p_1, \quad p_{0,i} = p_0, \quad i=1 \\ p_{1,i} = P_{e,i-1}^{OD}, \ p_{0,i} = 1 - p_{1,i}, \ i = 2,3 \cdots N \end{cases} \quad (1)$$

$$\frac{p_{0,i}(\beta_i - \xi\sqrt{\tau}\alpha_i)}{p_{1,i}(\beta_i + \xi\sqrt{\tau}\alpha_i)} = e^{-4\eta\xi\sqrt{\tau}\alpha_i\beta_i}, i=1,2,\cdots,N \quad (2)$$

$$P(C | \alpha_i, \beta_i^*) = p_{0,i} e^{-\nu-\eta(\tau|\alpha_i|^2 + |\beta_i^*|^2 - 2\xi\sqrt{\tau}\alpha_i\beta_i^*)} \\ + p_{1,i}[1 - e^{-\nu-\eta(\tau|\alpha_i|^2 + |\beta_i^*|^2 + 2\xi\sqrt{\tau}\alpha_i\beta_i^*)}] \quad (3)$$

$$P_{e,i}^{OD} = 1 - P(C | \alpha_i, \beta_i^*) \quad (4)$$

where $p_{q,i}$ ($q=0,1$) is *a priori* probabilities of hypothesis $H_{q,i}$ of each segment. $P_{e,i}^{OD}$ is the error probabilities of the ODR corresponding to each segment. The optimal displacement $\beta_i = \beta_i^*$ of each segment can be numerically solved from transcendental equation (2). $\eta$, $\nu$, $\tau$, and $\xi$ are quantum efficiency of detector, dark counts of detector, beam splitter transmittance and mode match factor of the ODR, respectively. At the end of the iteration process, the error probability of partitioned receiver with non-ideal factors $P_E$ is obtained,

$$P_E(t_1, t_2, \cdots t_N) = P_{e,N}^{OD} \quad (5)$$

## III. NUMERICAL SIMULATION RESULTS

From equation (5), with different partition strategies, the error probability is different. Varying $t_1, t_2, \cdots, t_N$ numerically, optimal partition strategy can be obtained by minimizing $P_E(t_1, t_2, \cdots, t_N)$. Vilnrotter's $N$-segment receiver works by iteratively considering the first ($N$–1) segments of an $N$-segment receiver as an optimized 2-segment receiver. In addition to Vilnrotter's partition strategy, another simple strategy is identical partition with equal $t_i$.

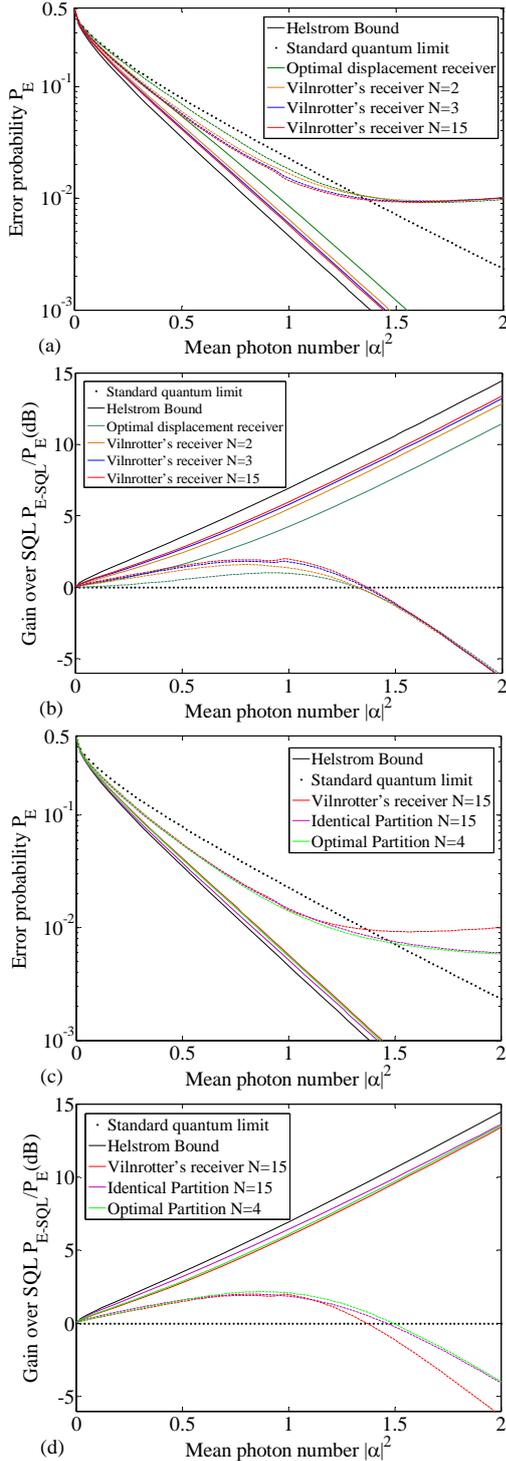

Fig. 2. The error probability (a), (c) and the gain over SQL (b), (d) of partitioned receiver. Solid line and dashed line corresponds to ideal case and non-ideal case, respectively. In both figures, the black solid lines and dotted lines represent the Helstrom bound and the SQL.

Fig. 2 illustrates the error probability and the gain over SQL of partitioned receiver. For ideal cases (solid lines), we use parameters $\eta = 1$, $v = 0$, $\tau = 1$, and $\xi = 1$, while for non-ideal cases (dashed lines), parameters $\eta = 0.9$, $v = 0.001$, $\tau = 0.99$, and $\xi = 0.995$ are used. In both figures, the black solid lines and dotted lines represent the Helstrom bound and the SQL. Form solid lines in fig. 2 (a) and (b), we note that in ideal cases, Vilnrotter's receiver works relatively better than SQL, but with increasing $N$, it rapidly approaches its performance limit. When the number $N$ is larger than 3, the additional performance gain is not obvious with increasing $N$. But in non-ideal conditions, as dashed lines in Fig. 2 (a) and (b), the situation is different. When the imperfect factors are considered, the receiver performance is degraded. In this situation, Vilnrotter's receiver gain over SQL becomes smaller than ideal case. What's more, for large mean photon number, the error probabilities are even high than SQL.

However, as seen from Fig. 2 (c) and (d), in ideal case, by simpler identical partition with large $N$, the receiver performance can surpass Vilnrotter's receiver performance limit. But in non-ideal cases, the gain is not obvious for small mean photon number.

In order to get more gain with small $N$ over SQL, global optimized partition strategy is used. As in Fig. 2 (c) and (d), though in ideal case, global optimized partition with small $N$ (here $N$=4) is not better than identical partition with large $N$ (here $N$=15), in non-ideal case, the former works slightly better than the latter.

## IV. CONCLUSIONS

According to the above results, global optimized partition with small $N$ is preferred for practical high rate implementation. It should be mentioned that, for some higher modulation formats, such as PPM and QPSK, sub-SQL quantum receiver have been experimentally demonstrated [7,8]. But physically realizable techniques for other modulation formats (such as QAM), and high rate implementation remains a major challenge.